\documentstyle[12pt]{article}
\begin{document} 
\global\parskip 6pt
\newcommand{\be}{\begin{equation}}
\newcommand{\ee}{\end{equation}}
\newcommand{\bea}{\begin{eqnarray}}
\newcommand{\eea}{\end{eqnarray}}
\newcommand{\non}{\nonumber}

\begin{titlepage}
\hfill{hep-th/0103108}
\vspace*{1cm}
\begin{center}
{\Large\bf The Cardy-Verlinde Formula and Taub-Bolt-AdS}\\
\vspace*{.2cm}
{\Large \bf Spacetimes}\\
\vspace*{2cm}
Danny Birmingham\footnote{E-mail: dannyb@pop3.ucd.ie}\\
\vspace*{.5cm}
{\em Department of Mathematical Physics\\
University College Dublin\\
Belfield, Dublin 4, Ireland}\\
\vspace*{1cm}
Soussan Mokhtari\footnote{E-mail: susan.mokhtari@itb.ie}\\
\vspace*{.5cm}
{\em Institute of Technology, Blanchardstown\\
Blanchardstown Road North\\
Dublin 15, Ireland}
\vspace{1cm}

\begin{abstract}
We consider the conformal field theory which is dual to the
Taub-Bolt-AdS spacetime. It is shown that the Cardy-Verlinde
formula for the entropy of the conformal field theory
agrees precisely with
the entropy of the Taub-Bolt-AdS spacetime, at high temperatures.
This result may be viewed as providing a conformal field theory
interpretation of Taub-Bolt-AdS entropy.
\end{abstract}
\vspace{.5cm}
March 2001
\end{center}
\end{titlepage}

In a recent paper \cite{verlinde1}, a proposal was put forward
for the entropy of a $D$-dimensional conformal field theory on
${\mathbf R} \times S^{D-1}$, with metric
\bea
ds^{2} = -dt^{2} + R^{2} d\Omega_{D-1}^{2}.
\label{cftmetric}
\eea
This Cardy-Verlinde formula takes the form
\bea
S_{\mathrm{CFT}} = \frac{2 \pi R}{D-1}\sqrt{E_{C}(2E - E_{C})},
\eea
and expresses the CFT entropy in terms of the energy $E$,
the Casimir energy $E_{C}$, and the radius $R$ of $S^{D-1}$.
The Casimir energy is defined \cite{verlinde1}
as the sub-extensive part of the energy
$E$.
For CFT in two dimensions, this formula reduces to the standard
Cardy formula. In higher dimensions, various aspects of this proposal
have been investigated \cite{verlinde1}-\cite{odintsov}.
In particular, the validity of the formula
has been established for the conformal field
theories which are dual to the Schwarzschild-AdS
black hole \cite{verlinde1} and Kerr-AdS black hole \cite{klemm}.
Other cases have been considered in \cite{cai}.

Within the context of the AdS/CFT correspondence \cite{Mald1}-\cite{W1},
one can study the boundary conformal field theory at finite temperature
defined on a manifold $S^{1} \times S^{D-1}$.
The relevant AdS configuration in this case is the
Euclidean section of the Schwarzschild-AdS black hole \cite{HP}.
It was shown in \cite{W2} that the entropy, energy,
and temperature of the boundary CFT can be identified
with the corresponding quantities in the black hole spacetime.
Furthermore, in the limit of high temperatures,
conformal invariance
can then be invoked to show that the Bekenstein-Hawking entropy
of the Schwarzschild-AdS black hole scales correctly with the horizon
volume. However, in order to fix the proportionality constant
between the entropy
and the horizon volume, one requires additional information, such as
a more detailed knowledge of the boundary conformal field theory.
The Cardy-Verlinde formula can be viewed as providing this additional
information.
In this way, one obtains a microscopic (CFT) interpretation of
black hole entropy. Alternatively, one can view the Schwarzschild-AdS
black hole as a testing ground for the verification of the
Cardy-Verlinde formula.

It is clearly of interest to study further examples of spacetimes
which have gravitational entropy. One such class is
the four-dimensional Taub-NUT-AdS and Taub-Bolt-AdS spacetimes
\cite{pope1}-\cite{page}.
These spacetimes have the property that they are only locally
asymptotically anti-de Sitter, and are parametrized by an integer $k$,
and a positive real number $s$.
In particular, the boundary CFT is then defined on a non-trivial
$S^{1}$ bundle over $S^{2}$.
The $k=1$ case is of particular interest, as it exhibits a phase
transition similar to the Schwarzschild-AdS case \cite{chamblin, page}.
Aspects of the dual CFT have been studied in \cite{dowker}.

In this paper, we present an explicit verification of the Cardy-Verlinde
formula for these Taub-NUT and Taub-Bolt spaces. We find
that the Cardy-Verlinde formula yields a CFT entropy in
precise agreement with the gravitational entropy,
in the limit of high temperature.
In \cite{chamblin,page}, the correct high temperature dependence
of the gravitational
entropy was shown to follow from the AdS/CFT correspondence.
Thus, the Cardy-Verlinde formula provides the additional
information necessary to fix the numerical coefficient in the
microscopic (CFT) derivation of the entropy of the Taub-Bolt-AdS spacetime.

The line element of the Taub-NUT-AdS metric can be
written in the form \cite{page}
\bea
ds^{2} &=& \frac{l^{2} B}{4}\left[\frac{F(r)}{B(r^{2} - 1)}(d\tau
+ B^{1/2}\cos\theta\;d\phi)^{2} + \frac{4(r^{2}-1)}{F(r)}dr^{2}
\right.\non\\
&+& \left.
(r^{2}-1)(d \theta^{2} + \sin^{2}\theta\;d\phi^{2})\right],
\label{nut}
\eea
where
\bea
F_{\mathrm{nut}}(r) = Br^{4} + (4-6B)r^{2} + (8B - 8)r +
(4 - 3B).
\eea
Here, $B$ is an arbitrary constant which is related to a `nut' charge,
and $\Lambda = -3/l^{2}$ is the cosmological constant.
The Euclidean time coordinate, $\tau$, has period $\beta = 4\pi B^{1/2}$.

The Taub-Bolt-AdS metric has the same form as (\ref{nut}),
with
\bea
F_{\mathrm{bolt}}(r)  = Br^{4} + (4 - 6B)r^{2} + \left(-Bs^{3} +
(6B - 4) s + \frac{3B-4}{s}\right)r + (4 - 3B),
\eea
and
\bea
B = \frac{2(ks - 2)}{3(s^{2} - 1)}.
\label{B}
\eea
Here, $k$ is the first Chern number of the $S^{1}$ bundle over
$S^{2}$, and $s$ is an arbitrary parameter, which must satisfy
the constraints $s > 1$, and $s > 2/k$. Also, $r > s$, and the
spacetime has a bolt at $r=s$.
The periodicity in Euclidean time is now $\beta = 4 \pi B^{1/2}/k$.
The parameter $s$ may lie on either an upper branch or lower branch,
defined by (\ref{B}), namely
\bea s_{\pm} = \frac{k}{3B}\left( 1 \pm \sqrt{1
- \frac{12 B}{k^{2}} + \frac{9B^{2}}{k^{2}}}\;\right).
\label{branch}
\eea

For our purposes here, we need only recall that the thermodynamics
of the Taub-NUT-AdS and Taub-Bolt-AdS spacetimes has been considered
in \cite{chamblin,page}. The crucial point is to decide on an
appropriate reference background in order to compute the action
of these spaces. An alternative approach without the use of
a background has been discussed in \cite{emparan, mann}.
In \cite{chamblin, page}, it was shown
that the relevant background is the
Taub-NUT-AdS spacetime with $k$ points identified
on $S^{1}$.
The action of the Taub-Bolt-AdS space relative to the Taub-NUT-AdS
background is then given by \cite{chamblin,page}
\bea
I = - \frac{\pi l^{2}}{18k} \frac{(ks - 2)[k(s^{2} + 2s + 3)
-4(2s + 1)]}{(s+1)^{2}}.
\eea
This leads directly to an expression for the entropy of the
Taub-Bolt-AdS space, via $S = \beta \partial_{\beta}I - I$.
This yields
\bea
S_{\mathrm{AdS}}
= \frac{\pi l^{2}}{6}\frac{(ks - 2)}{(s+1)^{2}}
\left[(s^{2} + 2s - 1) - \frac{4}{k}\right].
\label{entads}
\eea
Finally, the energy of the spacetime is given by $E = \partial_{\beta}I$,
namely
\bea
E = \frac{l^{2}}{36 B^{1/2}}\frac{(s-1)(ks-2)[k(s+3) + 4]}{(s+1)^{2}}.
\label{enerads}
\eea

As in the Schwarzschild-AdS case, the entropy and energy of
the strongly coupled dual
CFT can be identified with (\ref{entads}) and (\ref{enerads}).
In the present case, the three-dimensional
dual CFT is defined on a non-trivial $S^{1}$ bundle over $S^{2}$.
However, we should note that
the boundary metric (\ref{nut}) takes the form
\bea
\lim_{r \rightarrow \infty}\frac{4}{l^{2}B}\frac{R^{2}}{r^{2}}ds^{2} =
R^{2}(d\tau + B^{1/2}\cos \theta \;d\phi)^{2}
+ R^{2} (d\theta^{2} + \sin^{2}\theta\; d\phi^{2}).
\eea
Thus, in order to coincide with (\ref{cftmetric}),
the inverse temperature of the CFT
must be re-scaled by a factor of $R$ relative to the
inverse temperature of the Taub-Bolt-AdS spacetime.
Correspondingly,
the energy $E$ of the CFT is re-scaled by a factor of $1/R$
relative to (\ref{enerads}).
It is now straightforward to compute
the Casimir energy of the CFT as defined in \cite{verlinde1}, namely
\bea
E_{C} = DE - (D-1)TS.
\eea
Here, $\beta = 4 \pi B^{1/2}R/k$ is the inverse temperature of the CFT.
We find
\bea
E_{C} =  \frac{l^{2}}{6 R\;B^{1/2}}\frac{(ks-2)(2s - k)}{(s+1)^{2}}.
\label{ecas}
\eea
According to the Cardy-Verlinde formula, the entropy of the CFT
is then given by
\bea
S_{\mathrm{CFT}} &=& \pi\sqrt{E_{C}(2E - E_{C})}\non\\
&=& \frac{\pi l^{2}}{6} \frac{(ks-2)}{(s+1)^{2}}\left[\frac{1}{2}(2s-k)
(s+2)(s^{2} - 1)\right]^{1/2}.
\label{entcft}
\eea

It remains to check the relation between
this CFT entropy and the entropy of the AdS spacetime given by
(\ref{entads}).
In the limit of high temperature, $B \rightarrow 0$, we find that
the bolt size on the upper branch (\ref{branch}) takes the form
\bea
s_{+} = \frac{2k}{3B} - \frac{2}{k} + O(B).
\eea
In this limit, the energy and Casimir energy of the CFT are given by
\bea
E = \frac{l^{2}k^{3}}{54R\;B^{3/2}},\;\;E_{C} = \frac{l^{2}k}{3R\;B^{1/2}}.
\eea
As expected, the Casimir energy is smaller that the energy for high
temperature. Furthermore, we observe the expected leading order temperature
dependence, namely $E \sim T^{3}$, and $E_{C} \sim T$.
To leading order, the Cardy-Verlinde formula then
yields an entropy of the form
\bea
S_{\mathrm{CFT}} = \frac{\pi l^{2}}{6} ks =
\frac{\pi l^{2} k^{2}}{9B} = S_{\mathrm{AdS}}.
\eea
Thus, we see that the Cardy-Verlinde formula is in precise
numerical agreement
with the high-temperature limit of the gravitational entropy.
As predicted from the AdS/CFT correspondence,
the entropy scales as $\beta^{-2}$
\cite{chamblin, page}.
However, we now have the additional information required in order to fix
the numerical coefficient precisely.
Thus, one can state that the Cardy-Verlinde formula does indeed yield
the correct expression for the entropy of the CFT which is dual to
the Taub-Bolt-AdS
spacetime, at high temperature. However, the full functional form
given by (\ref{entads}) and (\ref{entcft}) do not agree at sub-leading order.
Of course, one should note
that in the limit of high temperature,
the entropy of the Taub-Bolt-AdS space reduces to the limit of
the standard formula $A_{\mathrm{bolt}}/4$, where the area of the
bolt is given by \cite{page}
$A_{\mathrm{bolt}} = (2 \pi l^{2}(ks - 2)/3)$.

It is worthwhile to consider the $k=1$ case more closely.
For temperatures above a critical value, there are two possible
Taub-Bolt-AdS spaces. The spacetime corresponding
to the larger value of $s$, discussed above, is thermodynamically
stable. However, the space with the smaller value
of $s$ is thermodynamically unstable \cite{chamblin, page}.
Nevertheless, it is useful to examine the Cardy-Verlinde formula
in this case.
In the limit of high temperature on the lower branch, we have
\bea
s_{-} = 2 + \frac{9B}{2} + O(B^{2}).
\eea
To leading order, the energy and Casimir energy of the CFT are
then given by
\bea E= \frac{l^{2}B^{1/2}}{8R},\;\;E_{C} = \frac{l^{2}B^{1/2}}{4R}.
\eea
On this branch, the Casimir energy is greater than the energy.
Furthermore, to leading order, the AdS entropy and CFT entropy
are given by
\bea
S_{\mathrm{AdS}} = \frac{\pi l^{2}B}{4},\;\;
S_{\mathrm{CFT}} = \frac{\pi l^{2}B}{2\sqrt{2}}.
\eea
Thus, we see that the numerical coefficients do not agree in this case.
We conclude that for the $k=1$ case, the Cardy-Verlinde formula
agrees with the entropy of the thermodynamically stable
Taub-Bolt-AdS spacetime, in the limit of high temperature.

In conclusion, we have verified the Cardy-Verlinde formula
for the entropy of CFT defined on non-trivial $S^{1}$ bundles
over $S^{2}$. This is achieved by considering the corresponding
dual spaces, which are given by the Taub-NUT-AdS and Taub-Bolt-AdS
metrics. In the limit of high temperature, we have shown
that indeed the Cardy-Verlinde formula for the CFT entropy
agrees precisely with the gravitational entropy.
Thus, we achieve both a CFT derivation of the Taub-Bolt-AdS entropy,
and a verification of the Cardy-Verlinde formula.

\end{document}